\documentclass[12pt,letterpaper]{article}
\usepackage[utf8]{inputenc}
\usepackage{amsmath,amssymb,mathtools}
\usepackage{lmodern}
\usepackage{placeins}
\usepackage[margin=1in]{geometry}
\usepackage{graphicx}
\usepackage{subcaption}

\DeclareMathOperator{\Tr}{Tr}

\def\be{\begin{equation}}
\def\ee{\end{equation}}
\def\bear{\begin{eqnarray}}
\def\eear{\end{eqnarray}}

\def\half{{{1\over 2}}}

\newcommand{\FIGUREDIRECTORY}{.}

\begin{document}

\begin{titlepage}
~
\vskip 0.5in
\begin{center}
{\Large
Entanglement entropy on a fuzzy sphere with a UV cutoff
\vskip 0.5in}
Hong Zhe Chen and Joanna L.~Karczmarek   
\vskip 0.3in
{\it
Department of Physics and Astronomy\\
University of British Columbia,
Vancouver, Canada V6T 1Z1}
\end{center}
\vskip 0.5in
\begin{abstract}
We introduce a UV cutoff into free scalar field theory on the 
noncommutative (fuzzy) two-sphere.  Due to the IR-UV connection,
varying the UV cutoff allows us to control the effective
nonlocality scale of the theory.
In the resulting fuzzy geometry, we establish which degrees
of freedom lie within a specific geometric subregion 
and compute the associated vacuum entanglement entropy.
Entanglement entropy for regions smaller than the 
effective nonlocality scale is extensive, while
entanglement entropy for regions larger than 
the effective nonlocality scale follows the area law.
This reproduces features previously obtained in the 
strong coupling regime through holography.  
We also show that mutual information is unaffected
by the UV cutoff.
\end{abstract}
\end{titlepage}

\section{Introduction}

As a tool used to investigate phenomena from
the nature of quantum criticality to the holographic origin
of space time, entanglement entropy in quantum field theories
has received wide interest.
One of the strengths of  geometric entanglement entropy (entanglement entropy associated
with a particular region of space) is that its definition is
independent of the details of the field theory, such as its degrees of freedom, and of the quantum state under consideration.
A reflection of this universality is the simple holographic interpretation
of geometric entanglement entropy as an area of a minimal surface \cite{Ryu:2006bv}.

Since entanglement entropy is universal, it is a good tool to study 
unconventional quantum field theories, such as those lacking in
locality.  Nonlocal theories have been argued to be
relevant to the study of holographic duals of flat space \cite{Li:2010dr},
while noncommutative theories in particular have been argued to be related to
black hole horizons and information scrambling 
\cite{Edalati:2012jj}.
Unsurprisingly, entanglement entropy in nonlocal theories can have drastically different
properties than it does in local theories.  For example,
the now very-well established `area law'---in the vacuum state,
the leading term of entanglement entropy grows (at most) proportionately
to the area of the boundary of the region of interest---
can fail if locality is not present.
Studying entanglement entropy in nonlocal theories can help us understand which
properties of entanglement entropy are dependent on locality and how.  Conversely,
detailed behaviour of entanglement entropy in nonlocal theories can illuminate
features of nonlocal theories such as the presence or absence of a UV-IR connection.

Vacuum entanglement entropy in noncommutative field theories has been studied both
holographically 
\cite{Barbon:2008sr,Barbon:2008ut,Fischler:2013gsa,Karczmarek:2013xxa} 
and directly \cite{Dou:2006ni,Karczmarek:2013jca,Sabella-Garnier:2014fda,
Sabella-Garnier:2017svs,Rabideau:2015via,Okuno:2015kuc,Suzuki:2016sca}. 
Holographically, for regions whose size is below a certain critical
length scale $L_{\mathrm{trans}}$, the entanglement entropy grows with the volume of the region,
while the area law is restored for regions whose 
dimensions are larger than the critical length scale $L_{\mathrm{trans}}$.  
The critical length scale $L_{\mathrm{trans}}$ is proportional to the UV cutoff,
revealing information about the
UV-IR connection.  It is not immediately clear whether the holographic
approach is truly measuring geometric entanglement entropy in the
field theory, as existence of well defined
geometric regions in the nonlocal boundary theory corresponding to 
regions of AdS boundary is not immediately obvious.
To remove these doubts, entanglement entropy was studied for 
a free field on a fuzzy (noncommutative) sphere directly, using a finite $N$
matrix model \cite{Karczmarek:2013jca}.
However, owing to an inevitable
relationship between the noncommutativity parameter 
$\sqrt{\boldsymbol{\theta}}$, 
the circumference of the sphere $R$ and the UV cutoff $\lambda_{\mathrm{UV}}$:
\be
2 R^2 = N {\boldsymbol{\theta}}~~~\mathrm{and}~~~\lambda_{\mathrm{UV}} = R/N~,
\ee
the expected critical length scale 
$L_{\mathrm{trans}} \sim {\boldsymbol{\theta}} / \lambda_{\mathrm{UV}} $ 
is larger than the radius of the sphere, making
it impossible to examine regions whose dimensions larger 
than $L_{\mathrm{trans}}$.

\begin{figure}
\includegraphics[width=\textwidth]{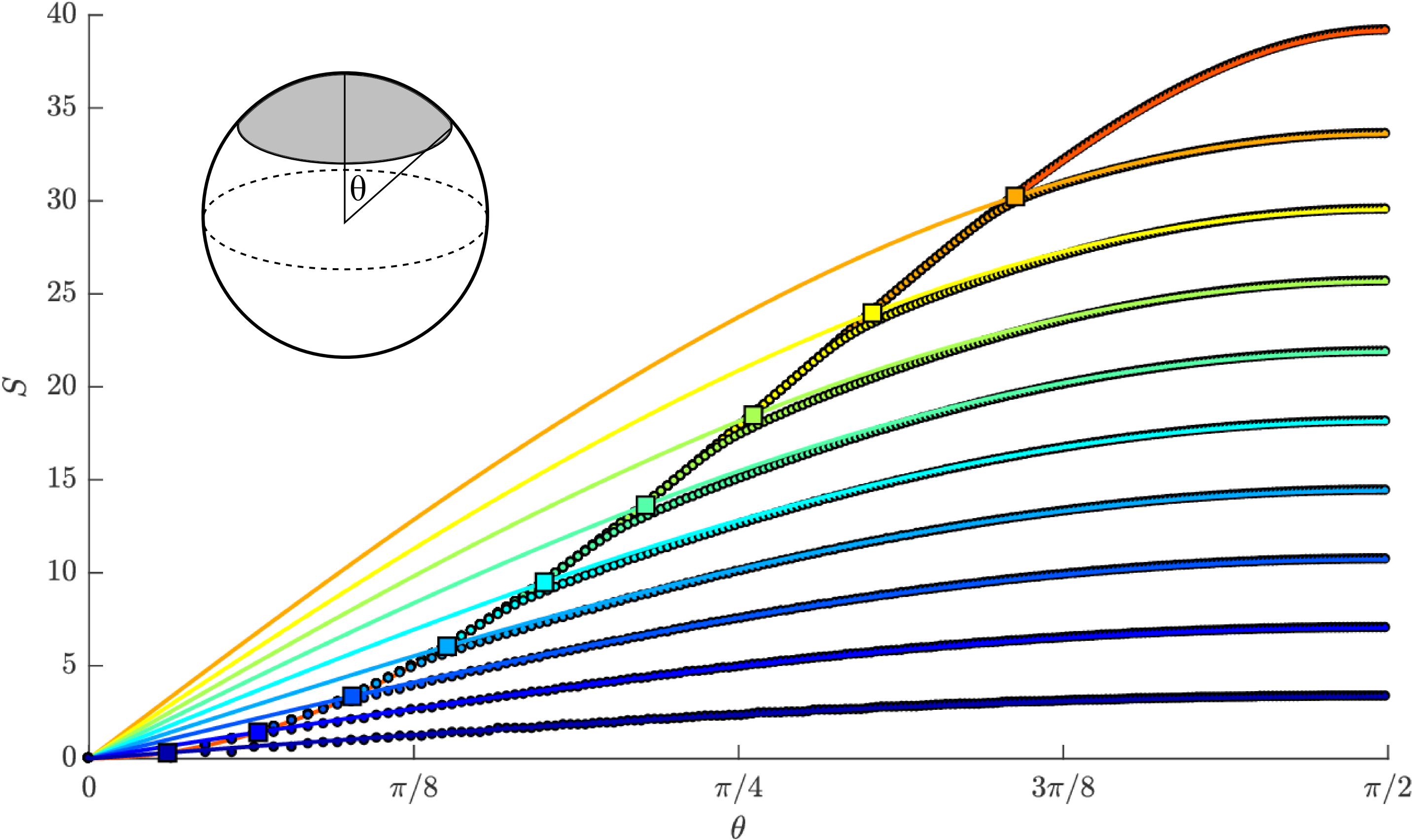} 
\caption{Entanglement entropy associated with a polar cap region
against cap's angular size $\theta$ for various UV cutoffs. 
Here, $N=200$, $\mu=1$, and $n$ takes values $20,40,\ldots,N$, indicated
with colors from dark blue ($n=20$) to redish orange ($n=N=200$). For $n<N$, solid
lines show $S(\theta=\pi/2)\sin(\theta)$,
which is proportional to the length of the cap's boundary.}
\label{S_vs_theta_sine_curves}
\end{figure}

In this paper, we give a prescription for lowering the UV cutoff
in a scalar field theory on a fuzzy sphere, 
increasing the associated wavelength 
from $\lambda_{\mathrm{UV}} = R/N$ to $\tilde \lambda_{\mathrm{UV}} = R/n$,
and defining geometric regions in a theory with a UV cutoff.   
In this simple theory, we are able to recover the phase transition 
previously seen holographically, as described above.
As can be seen in Figure \ref{S_vs_theta_sine_curves}, imposing a
lower UV cutoff has no effect on entanglement entropy associated 
with a spherical cap of angular size $\theta$ when $\theta$ is small.
For larger regions, entanglement
entropy becomes proportional to the length of the boundary of the
region, $2\pi R \sin \theta$.    As expected,
this transition from extensive entanglement entropy to area law
happens at an angular size $\theta_\mathrm{trans}$ which increases
proportionately to $1/\tilde \lambda_{\mathrm{UV}} \sim n$ (Figure \ref{vary_n}).

The rest of this paper is organized as follows.  
We work with a free scalar field theory on a noncommutative---or fuzzy---sphere
with $R=1$ throughout the paper.
In section
\ref{sec:method} we describe our approach to computing entanglement
entropy associated with a spherical cap region in the presence of
a UV cutoff.  In section \ref{sec:ee} we present detailed
behaviour of the entanglement entropy as we vary $\theta$, $n$, $N$ and the
mass of the scalar field.  In section \ref{sec:mutual}, we
discuss  mutual information.

\section{Methodology}
\label{sec:method}

The Hamiltonian for a real scalar field theory on a fuzzy sphere
of radius one is given by
\be
H = \frac{4 \pi}{N}
\half \Tr \left ( (\dot \phi)^2 ~-~ \sum_{i=1,2,3} [L_i,\phi]^2 ~+~ \mu^2  \phi^2 \right )
~,
\label{eq:H}
\ee
where $\phi$ is an $N\times N$ Hermitian matrix representing the
scalar field, whose mass is $\mu$. 
$L_i$ generate an irreducible
representation of the SU(2) algebra:
\be
[L_i, L_j] = \sum_k i\epsilon_{ijk} L_k~,~~~\sum_{i=1,2,3} L_i^2 = (N^2-1)/4~.
\ee
The Hamiltonian (\ref{eq:H}) represents $N^2$ coupled harmonic oscillators
of equal masses.  We work in a basis where $L_3$ is diagonal.
In \cite{Dou:2006ni} it was noticed that Hamiltonian (\ref{eq:H}) is a 
sum of $2N+1$ independent (mutually non-interacting) sectors.
This decomposition can be viewed as a consequence of rotational
symmetry.  To implement it, we need to first `complexify' the 
scalar field $\phi$.  The Hermitian matrix $\phi$ represents the configuration
space of our system, $\mathbb{R}^{N^2}$.  First, let's repackage
the information contained in $\phi$ to make its nature as a real
vector space easier to see, by defining a real matrix
$\tilde \phi = \half (\phi + \phi^T) + \frac{1}{2i}(\phi - \phi^T)$.
Matrices $\phi$ and $\tilde \phi$ are in one-to-one correspondence
with each other, with $\phi = \half(\tilde \phi + \tilde\phi^T) +
\frac{i}{2}(\tilde \phi - \tilde \phi^T)$, and $\tilde \phi$ is
a real matrix iff $\phi$ is Hermitian, thus each entry in matrix $\tilde \phi$
is an independent real field.
 The Hamiltonian (\ref{eq:H}) can
be rewritten in terms of $\tilde \phi$,
\be
H = \frac{4 \pi}{N}
\half \Tr \left ( \dot {\tilde \phi}^T\dot {\tilde \phi }
~-~ \sum_{i=1,2,3} [L_i,\tilde\phi^T][L_i,\tilde\phi] ~+~ 
\mu^2  \tilde \phi^T\tilde \phi 
\right )~, 
\label{eq:Htilde}
\ee
while the U(N) invariant metric (from which the measure
derives) on configuration space is given by
\be
ds^2 = \Tr d\phi^2  = \Tr d\tilde \phi^T d\tilde \phi~.
\ee
It is now easy to see that the Hamiltonian (\ref{eq:Htilde})
contains $2N+1$ independent sectors.
In (\ref{eq:Htilde}), consider $\tilde \phi$
to be a complex matrix and replace transpose with Hermitian
conjugate.  On real matrices $\tilde \phi$, the Hamiltonian
is unchanged.  Now, consider the action of $e^{i\alpha L_3}$
on the complexified $\tilde \phi$: $\tilde \phi \rightarrow
e^{i\alpha L_3} \tilde \phi e^{-i\alpha L_3}$. It is easy to see
that the Hamiltonian is unchanged under this action.  
Since under this action each term $\tilde \phi_{ab}$
in the complexified $\tilde \phi$ acquires a phase $e^{i(a-b)\alpha}$,
we learn that the Hamiltonian contains terms with 
$(\tilde \phi^\dagger)_{dc}\tilde \phi_{ab}=
\tilde \phi_{cd}^\star \tilde \phi_{ab}$ only if $(a-b)-(c-d)$
is zero.  Reducing $\tilde \phi$ back to a real matrix,
we see that the Hamiltonian does indeed contain $2N+1$ independent
sectors $V^{(m)}$, each consisting of terms $\tilde \phi_{ab}$ with
a fixed $a-b=m$, for $m$ from $-N$ to $N$.  We are able to compute entanglement
entropy in each of those sectors separately, greatly speeding
up the numerical analysis.

In each sector $V^{(m)}$, the Laplacian 
 operator 
 $\Delta : \tilde\phi \rightarrow \sum_{i=1,2,3} [L_i,[L_i,\tilde\phi]]$
has the non-degenerate eigenvalues
$j(j+1)$ with corresponding eigenvectors $v^{(m)}_j$
for $j$ from $|m|$ to $N-1$.
To impose a UV cutoff, we focus on a subspace of $V^{(m)}$ spanned
by the  $n-|m|$ eigenvectors $v^{(m)}_j$
for $j$ from $|m|$ to $n-1$.
Denote such a subspace with $V^{(m)}_n$ and let
 $O^{(m)}_n: V^{(m)} \rightarrow V^{(m)}_n$
be a linear operator such that
$O^{(m)}_n(v^{(m)}_j) = v^{(m)}_j$ for $j\in\{|m|, \ldots, n-1\}$
and $O^{(m)}_n(v^{(m)}_j) = 0$ for $j\in\{n, \ldots, N-1\}$.
Under the natural inner product, the eigenvectors of the 
Laplacian are orthogonal, so we also have the transpose
of $O^{(m)}_n$, $(O^{(m)}_n)^T: V^{(m)}_n \rightarrow V^{(m)}$
with  $(O^{(m)}_n)^T(v^{(m)}_j) = v^{(m)}_j$ for $j\in\{|m|, \ldots, n-1\}$.
(We use a transpose rather than 
a Hermitian conjugate as $V^{(m)}$ and $V^{(m)}_n$ are real vector spaces.)

In \cite{Karczmarek:2013jca} it was argued that the degrees of 
freedom associated with a spherical cap centered on the north pole
are represented by matrix entries $\phi_{ab}$ 
(or $\tilde \phi_{ab}$) with $a+b < N (1-\cos(\theta))$.
For maximal control over
selection of degrees of freedom, we define an operator $Z$,
which `measures' the position of any degree of freedom along the
$L_3$ (or $z$) axis.  This operator acts on any matrix entry
$\phi_{ab}$ by multiplying it by $N+1-(a+b)$.  When $Z$ is restricted to
a single sector $V^{(m)}$, we denote the corresponding
operator $Z^{(m)}$.  
Thus, we can easily identify, given $\theta$, which degrees of freedom in
each sector $V^{(m)}$ are inside or outside the spherical cap.
We denote with $P_{\theta}^{(m)}$ the operator acting on
a sector $V^{(m)}$ which projects onto the degrees
of freedom within a given cap (with $(P^{(m)}_{\theta})^2=
P^{(m)}_{\theta}$).  

The UV cutoff is associated with a projection operator
$(O^{(m)}_n)^T (O^{(m)}_n)$ which does not commute with
$P_{\theta}^{(m)}$ or $Z^{(m)}$.
This is not surprising, and implies that imposing a cutoff `blurs' the edges
of the spherical cap region.
To consider a spherical cap in a theory with a UV
cutoff, we can replace $P^{(m)}_{\theta}$ with
$P^{(m)}_{\theta,n}:=O_n^{(m)} P_{\theta}^{(m)}  
(O_n^{(m)})^T$, which acts on $V^{(m)}_n$.
$P^{(m)}_{\theta,n}$ is not a projection operator itself.
In a UV-restricted theory,
we would identify degrees of freedom inside
the spherical cap with an eigenspace of $P^{(m)}_{\theta,n}$ whose
eigenvalues are greater than half.
This is sensible as the eigenvalues of $P^{(m)}_{\theta,n}$ are between $0$ and $1$ by
construction, with most of them close to either $0$ or $1$ and only
a small subset falling in-between.  

A better (it turns out) method to assign the degrees of freedom uses
the operator $Z^{(m)}_n:=O_n^{(m)} Z^{(m)}  (O_n^{(m)})^T$.
The inside of the spherical cap is associated with eigenvectors
of $Z^{(m)}_n$ whose eigenvalues are greater than $z=(N-\half)\cos\theta$.
While one might think naively that methods based on operators
$P^{(m)}_{\theta,n}$ and $Z^{(m)}_n$ are
interchangeble (at least in the large $N$ limit), this turns
out to not be the case.  While for entanglement entropy, the two
methods do produce similar (though not identical) results, 
significant artifacts are introduced into mutual information if the operator
$P^{(m)}_{\theta,n}$ is used to assign degrees of freedom belonging to a 
region (see section \ref{sec:mutual}). 
This highlights the delicate nature of delineating geometric
regions in a fuzzy geometry.

All the computations presented in the paper were
obtained using $Z_n^{(m)}$.
In each UV-controlled sector $V^{(m)}_n$, we have two 
self-adjoint operators, the Laplacian $\Delta$ and $Z^{(m)}_n$.  If we write
the Hamiltonian in the basis in which $Z^{(m)}_n$ is diagonal,
we have a system of $n-|m|$ coupled harmonic oscillators.
We can use well-established procedures \cite{Bombelli:1986rw,Srednicki:1993im}
 for computing the entanglement
entropy of a selected number of those oscillators (using the above
criteria for which  degrees of freedom fall inside/outside
a given spherical cap).

\begin{figure}
\begin{subfigure}[t]{\textwidth}
\includegraphics[width=\textwidth]{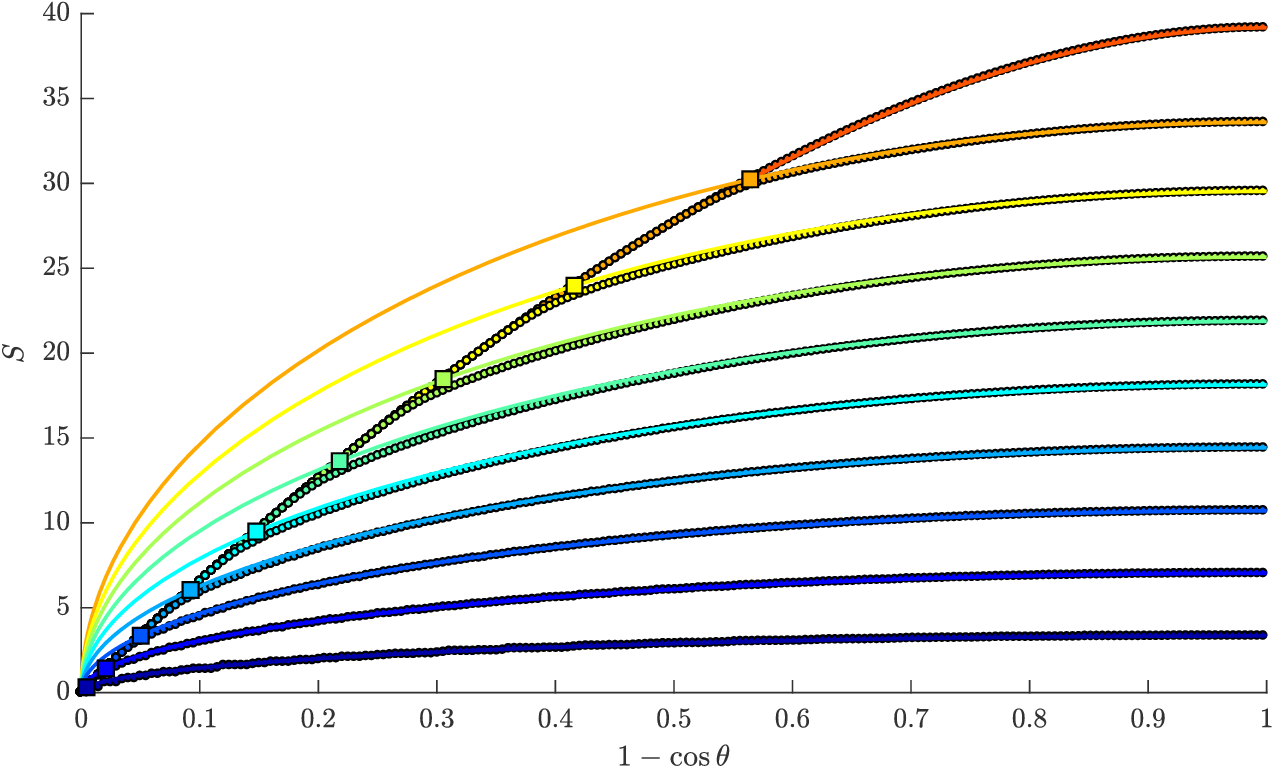}
\label{S_vs_trigfuncs_a}
\end{subfigure}
\begin{subfigure}[t]{\textwidth}
\includegraphics[width=\textwidth]{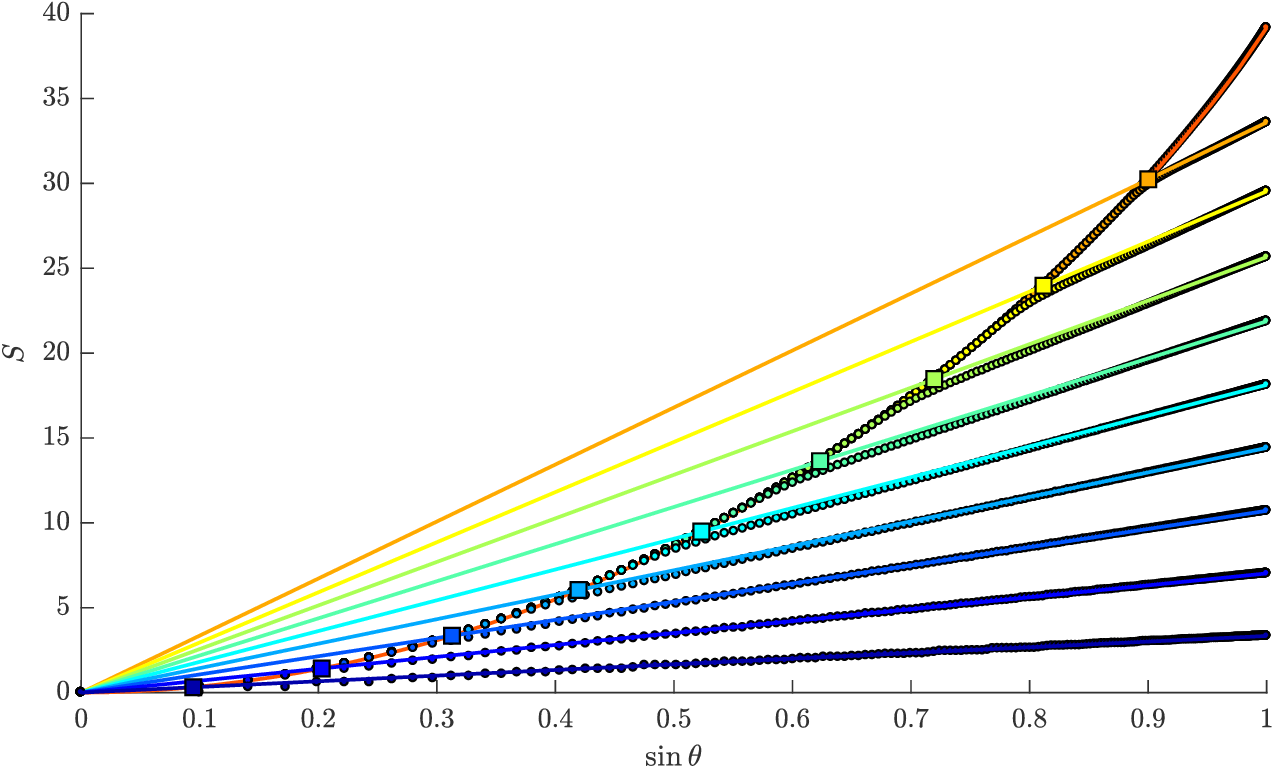}
\label{S_vs_trigfuncs_b}
\end{subfigure}
\caption{Entanglement entropy $S$ of a polar cap plotted against 
$1-\cos\theta$ (proportional to the area of the polar cap) and 
$\sin\theta$ (proportional to the length of the polar cap boundary) 
for $N=200$, for the data shown in Figure \ref{S_vs_theta_sine_curves}.}
\label{S_vs_trigfuncs}
\end{figure}

\section{Entanglement entropy}
\label{sec:ee}

Leading order entanglement entropy of a spherical
cap region on a commutative sphere is 
proportional to $n \sin\theta$, where $n$ is 
a discretization parameter such that the total
number of degrees of freedom is proportional to $n^2$
\cite{Sabella-Garnier:2014fda}.
We can think of the discretized commutative sphere
as a noncommutative sphere with  $N$ taken to infinity while
$n$ is held fixed.  We introduce a notation, $S(N,n;\theta)$
for the entanglement entropy of a spherical cap in a 
noncommutative sphere; $S(\infty,n;\theta) = \mathrm{const}\cdot n\sin\theta$
is then the local (commutative) result.

In Figure \ref{S_vs_theta_sine_curves}, entanglement entropy 
in noncommutative field theory
is plotted against cap 
size for a selection of UV cutoff parameters $n$. 
For each value of $n<N$, a sine curve is plotted at an 
amplitude matching the entanglement entropy at $\theta=\pi/2$. 
It is clear that imposing a UV cutoff has no effect on entanglement
entropy for small cap sizes, and that entanglement entropy becomes
proportional to the length of the cap's boundary once the cap
is large enough at a given UV cutoff.  This is further highlighted in
Figure \ref{S_vs_trigfuncs}, where the entanglement entropy is plotted
as a function of the cap's area (top) and as a function of the length of the
cap's boundary (bottom).  At any given UV cutoff, as $\theta$ grows, 
there is a transition from extensive behaviour characteristic of 
a highly non-local theory to area-law behaviour characteristic of
a local theory.

\begin{figure}
\includegraphics[width=\textwidth]{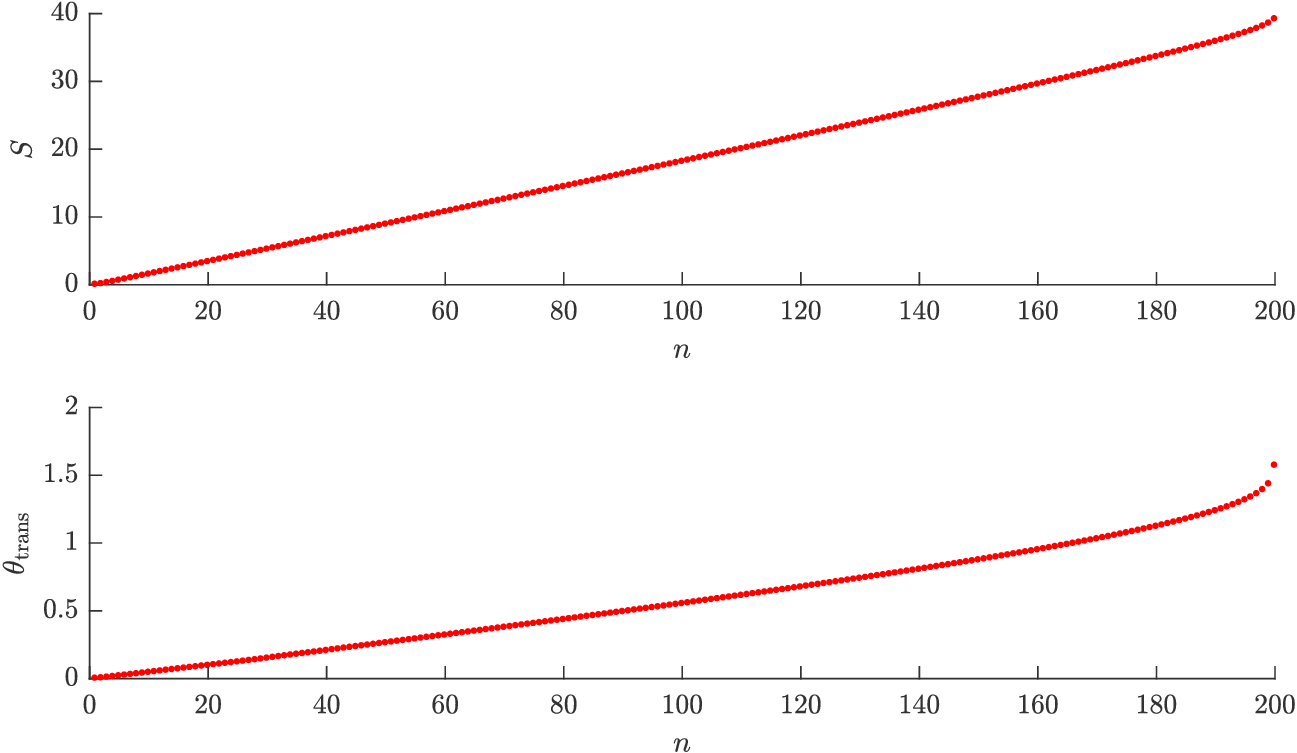}
\caption{Entanglement entropy of a hemisphere (top) and 
transition angle $\theta_\mathrm{trans}$ from 
Figure \ref{S_vs_theta_sine_curves}
(bottom) against the UV cutoff parameter $n$. 
$N=200$, $\mu=1$.}
\label{vary_n}
\end{figure}

In Figures  \ref{S_vs_theta_sine_curves} and \ref{S_vs_trigfuncs}, 
we highlight the point at which this transition happens with colored squares.
Since this is a smooth cross-over rather than a sharp phase transition, the
exact position of the cross-over is not well defined; we choose to
use the intersection between sine curves (which represent behaviour expected 
from a local theory) and the interpolated curve for $n=N$ (which represents 
nonlocal behaviour).  We have computed this cross-over point as a function
of the cutoff parameter $n$; its roughly linear behaviour with $n$ can be seen in Figure
\ref{vary_n} (bottom).

This linear behaviour of the transition point between non-local and local
behaviour in entanglement entropy is a clear demonstration of the UV-IR
connection on the fuzzy sphere.  The UV-IR connection in noncommutative
 theories \cite{Minwalla:1999px} implies that for a theory with a noncommutativity
 scale $\sqrt{\boldsymbol{\theta}}$ defined by 
$[x_1,x_2]=i{\boldsymbol{\theta}}$,
the nonlocality scale is ${\boldsymbol{\theta}}/\lambda_{\mathrm{UV}}$,
where $\lambda_{\mathrm{UV}}$ is the UV cutoff wavelength.  On the fuzzy
sphere with radius $R$, we have ${\boldsymbol{\theta}} = R^2/J$ 
(where $N=2J+1$),
and the UV cutoff is $\tilde \lambda_{\mathrm{UV}} = R/n$, implying that
the transition should happen at $\theta_\mathrm{trans} = \mathrm{const}\cdot n/N$,
consistent with the behaviour shown in Figure \ref{vary_n}.

\FloatBarrier

\begin{figure}
\includegraphics[width=\textwidth]{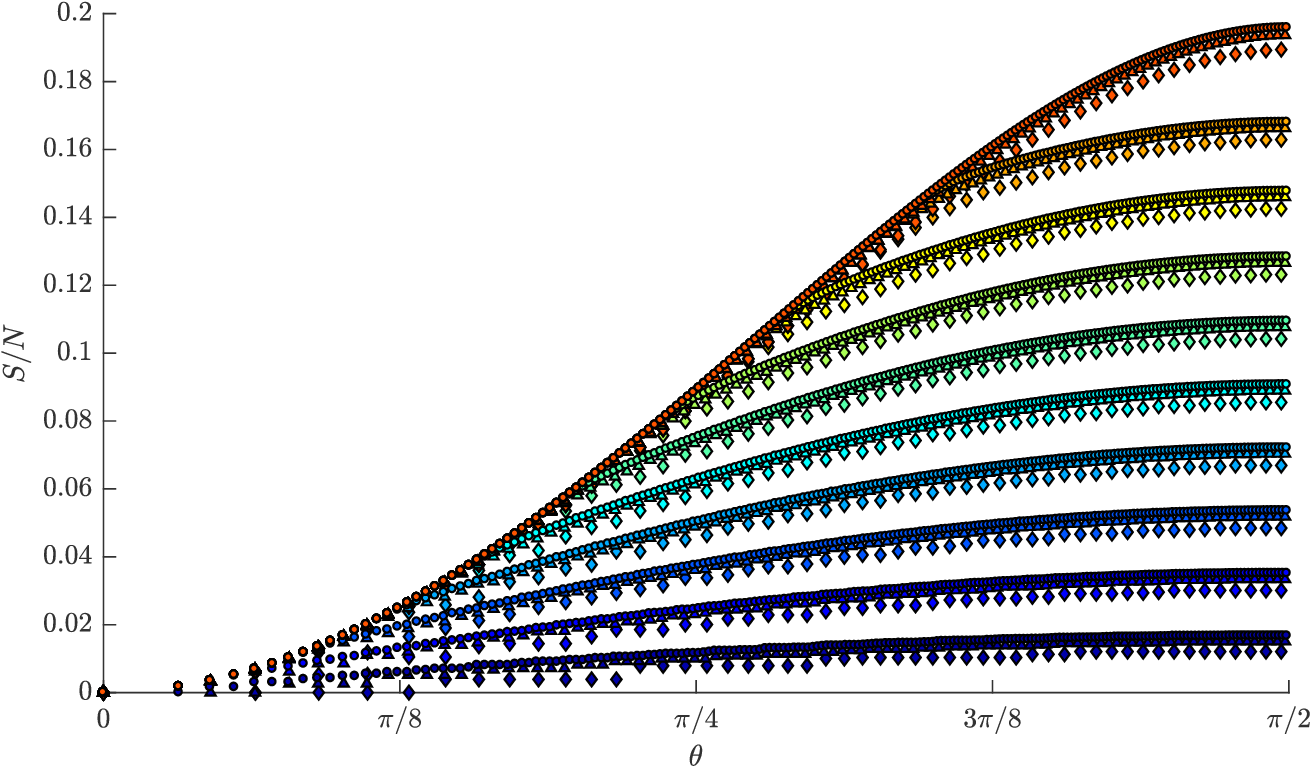}
\caption{Entanglement entropy divided by the quantization parameter,
$S/N$, plotted against angular cap size $\theta$ 
for various UV cutoffs and various $N$. 
The large diamonds, medium triangles, and small circles indicate 
$N=50,100,200$ respectively.  The UV cutoff parameters are proportional
to $N$, and given by $n=N/10, 2N/10, \ldots, N$, indicated with
colors from dark blue to redish orange. }
\label{vary_N}
\end{figure}

To examine the behaviour of entanglement entropy in more detail, 
we begin with dependence on the quantization parameter $N$.
As was already observed in \cite{Karczmarek:2013jca}, entanglement entropy
$S$ for cap with a fixed angular size grows linearly with $N$.  
This is peculiar in a nonlocal theory with 
$N^2$ degrees of freedom: we could naively expect entanglement entropy to
be proportional to $N^2$.  However, the theory's nonlocality is not
complete, allowing it to reproduce the behaviour of the local theory
when expected.
In Figure \ref{vary_N}, we plot $S(N,n;\theta)/N$ as a function of $\theta$,
to show that it converges to a function of $n/N$ and $\theta$.

Further, at a fixed $N$, the entanglement entropy of a hemisphere,
$S(N,n,\pi/2)$, is proportional to the cutoff parameter $n$,
as can be seen in Figure \ref{vary_n} (top). 
Together, Figures \ref{vary_n} and \ref{vary_N} establish that
the entanglement entropy of any spherical cap with $\theta>\theta_{\mathrm{trans}}$
is proportional to the UV cutoff parameter $n$.  

Entanglement entropy for small regions 
is smaller than the local theory would predict.
Further, for small regions, the entanglement entropy is independent
of the imposed UV cutoff (though it is still proportional to $N$).
This is a sign that only low-momentum modes participate in entanglement
entropy when $\theta < \theta_\mathrm{trans}$.

In summary, in the large $N$ limit, we can write
\be
S(N,n;\theta) = 
\begin{cases}
N s_{\mathrm{extensive}}(\theta) 
&~~~\mathrm{for}~\theta<\theta_\mathrm{trans}(n/N) \\
C n \sin(\theta) 
&~~~\mathrm{for}~\theta>\theta_\mathrm{trans}(n/N) 
\end{cases}~,
\ee
where $C$ is a constant,  
$\theta_\mathrm{trans}$ is a function of $n/N$ only,
and $s_{\mathrm{extensive}}(\theta) \sim \theta^2$ for
small $\theta$.
The above result includes behaviour of a commutative theory,
$S(\infty,n;\theta)=C n\sin(\theta)$, since for $n/N=0$, 
$\theta_\mathrm{trans}=0$.

\begin{figure}[h]
\includegraphics[width=\textwidth]{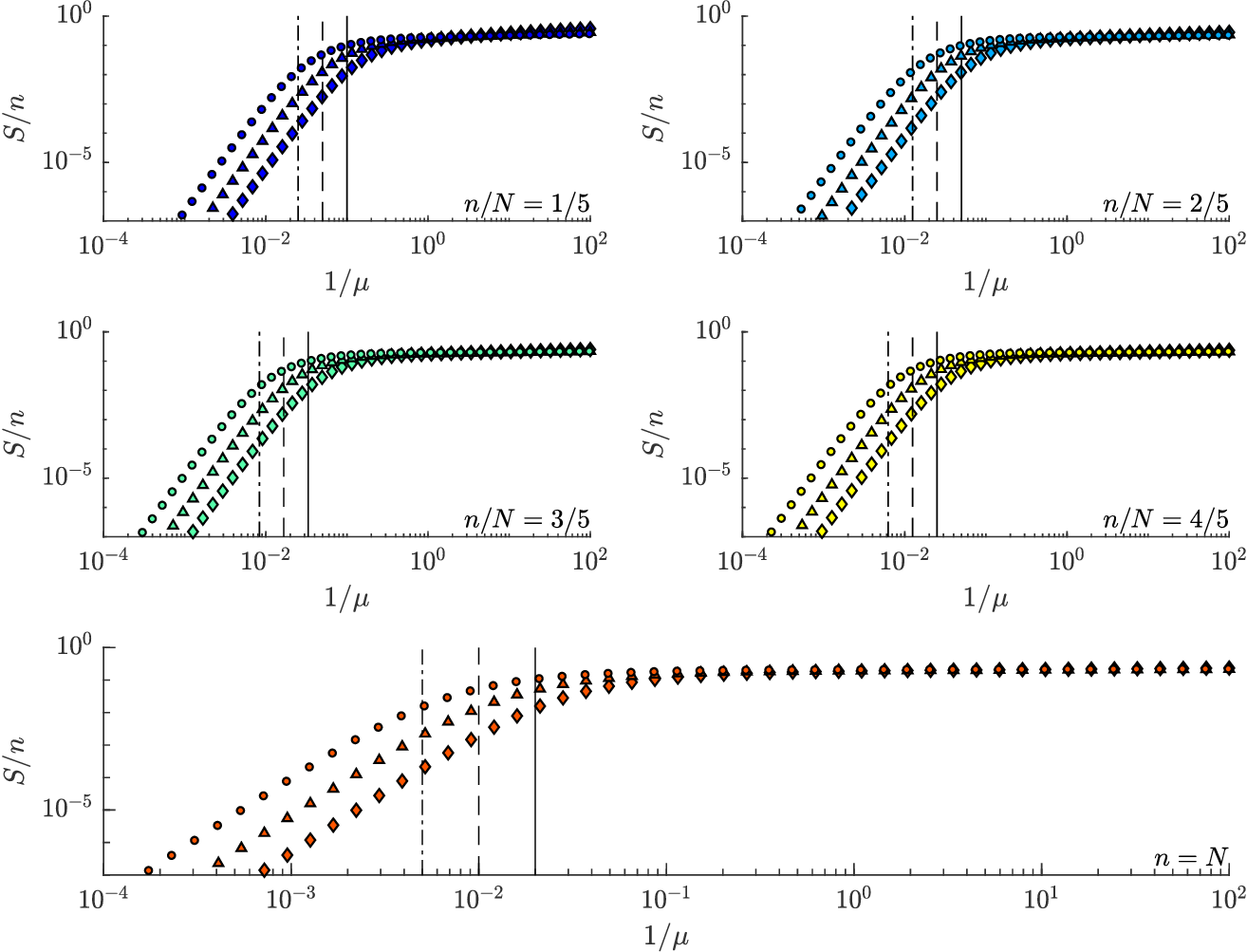}
\caption{Scaled entropy $S/n$ of a hemisphere plotted against $1/\mu$ for various UV cutoffs and various $N$. The UV cutoffs correspond to $n=N/5, 2N/5, \ldots, N$, as indicated. 
Large diamonds, medium triangles, and small circles indicate $N=50$, 
$100$ and $200$, respectively. Corresponding to those values of $N$, 
the solid, dashed, and dash-dotted lines mark $\mu=n$.}
\label{vary_mu}
\end{figure}

\begin{figure}
\begin{subfigure}[t]{\textwidth}
\includegraphics[width=\textwidth]{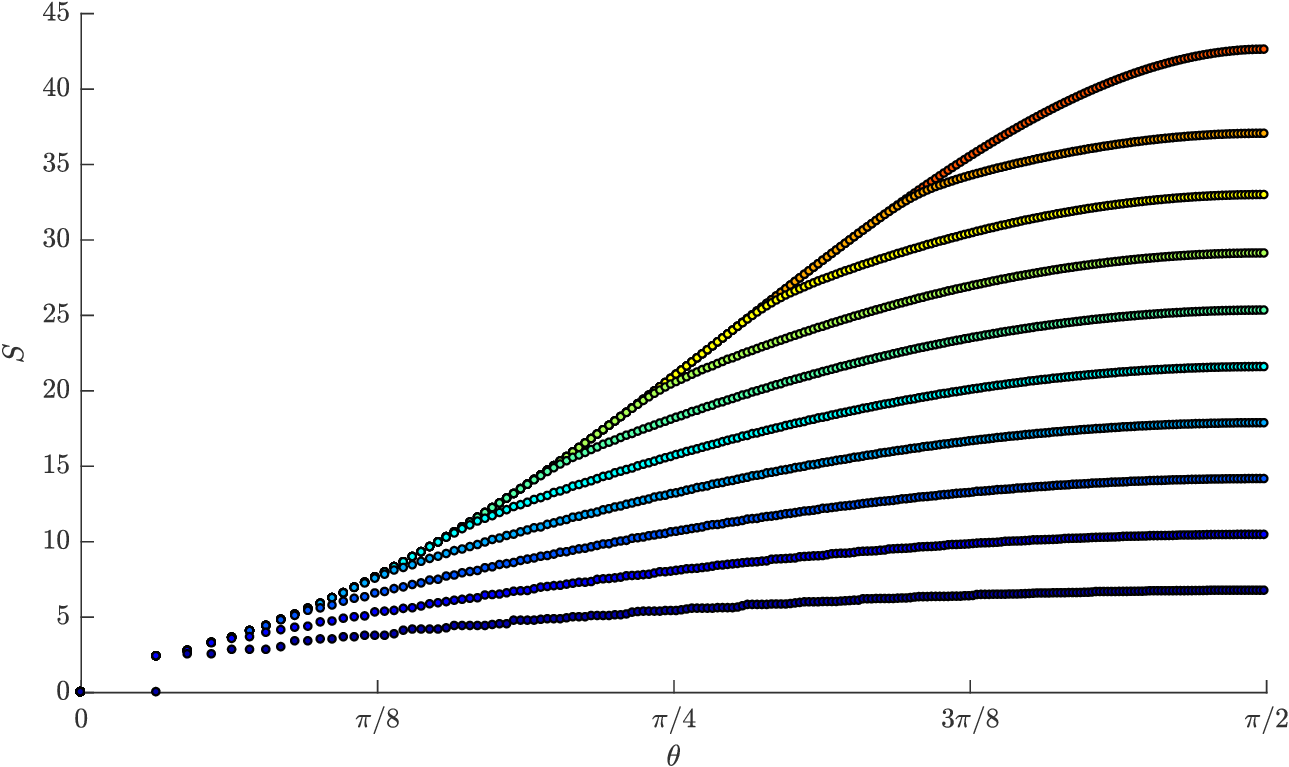}
\caption{$\mu=1/1000$}
\label{S_vs_theta_N200_mu1E-3}
\end{subfigure}
\begin{subfigure}[t]{\textwidth}
\includegraphics[width=\textwidth]{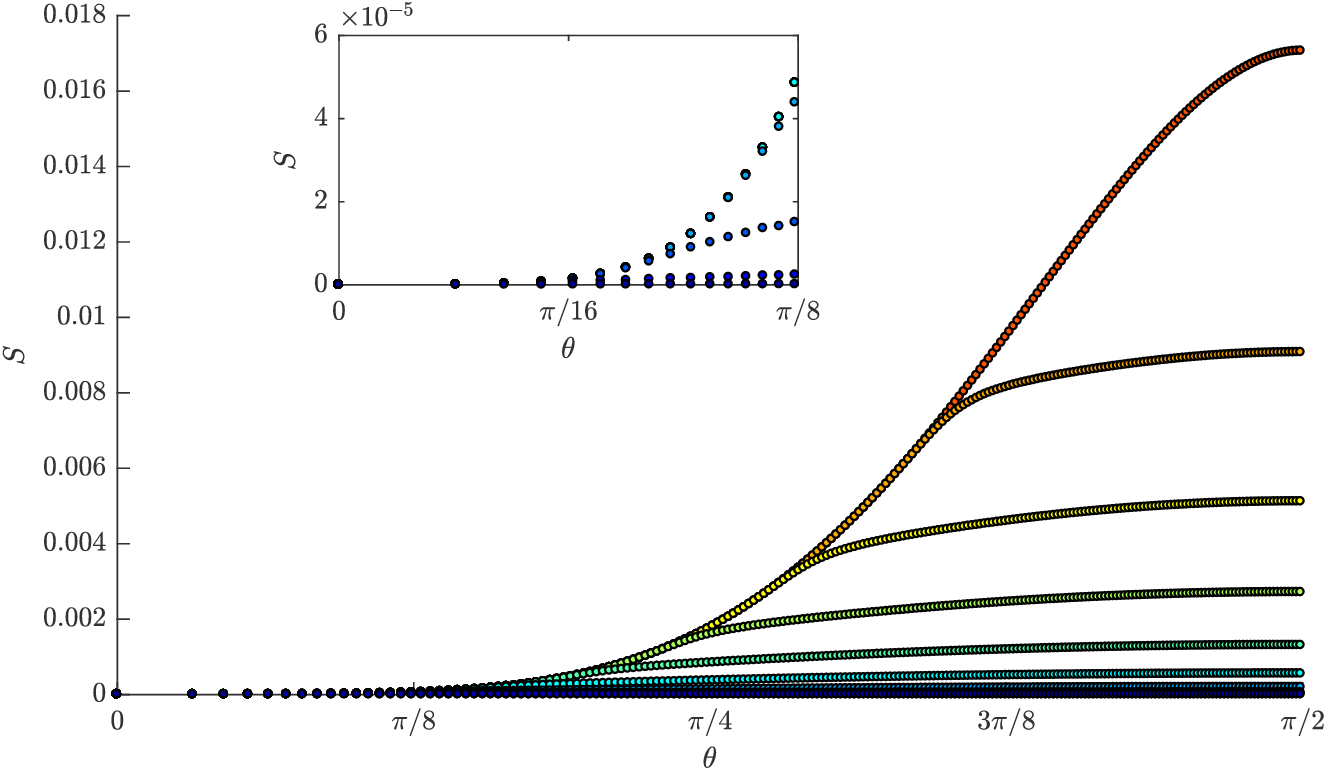}
\caption{$\mu=1000$}
\label{S_vs_theta_N200_mu1E3}
\end{subfigure}
\caption{Entropy $S$ of a spherical cap as a function of 
$\theta$ for $N=200$ with (a) a small mass $\mu=1/1000$ and (b) a 
large mass $\mu=1000$.  
UV cutoff parameters are $n=20,40,\ldots,200$ indicated with colors
from dark blue to redish orange.}
\label{S_vs_theta_N200_mu}
\end{figure}

\begin{figure}
    \centering
    \begin{subfigure}[t]{0.49\textwidth}
    \centering
    \includegraphics[width=\textwidth]{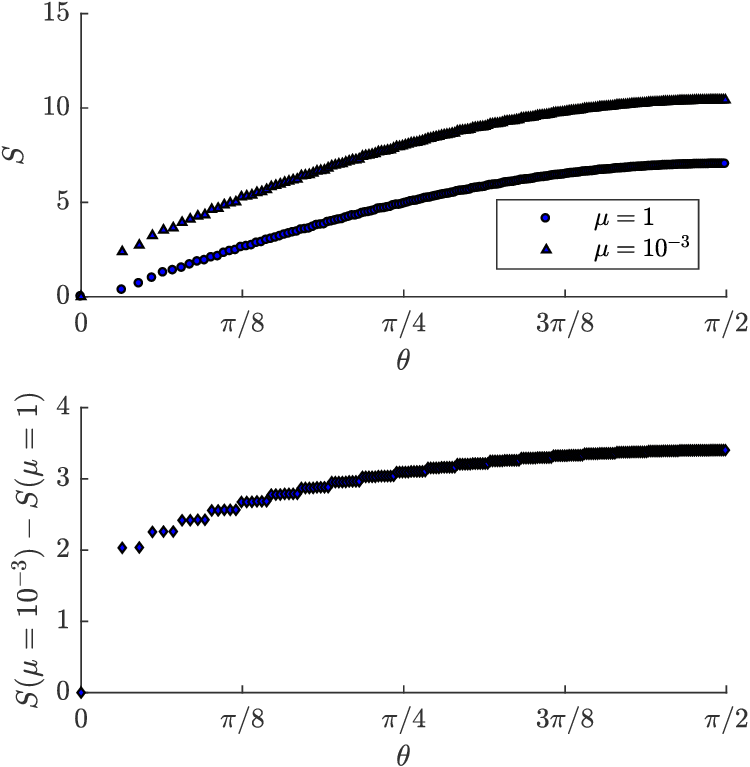}
    \caption{$n/N=1/5$}
    \label{S_vs_theta_beta1over5_mu1mu1E-3}
    \end{subfigure}
    \begin{subfigure}[t]{0.49\textwidth}
    \centering
    \includegraphics[width=\textwidth]{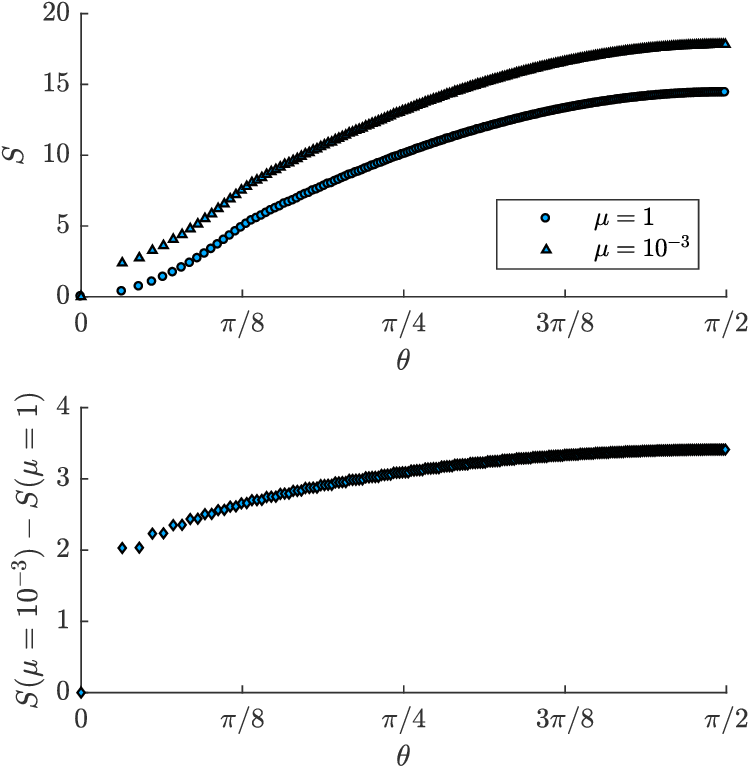}
    \caption{$n/N=2/5$}
    \label{S_vs_theta_beta2over5_mu1mu1E-3}
    \end{subfigure}
    \begin{subfigure}[t]{0.49\textwidth}
    \centering
    \includegraphics[width=\textwidth]{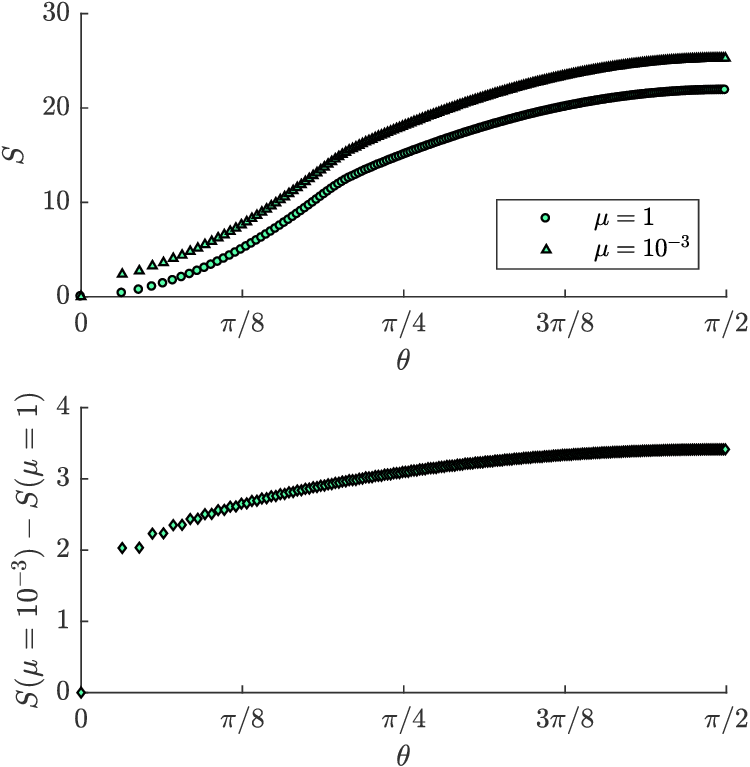}
    \caption{$ n/N=3/5$}
    \label{S_vs_theta_beta3over5_mu1mu1E-3}
    \end{subfigure}
    \begin{subfigure}[t]{0.49\textwidth}
    \centering
    \includegraphics[width=\textwidth]{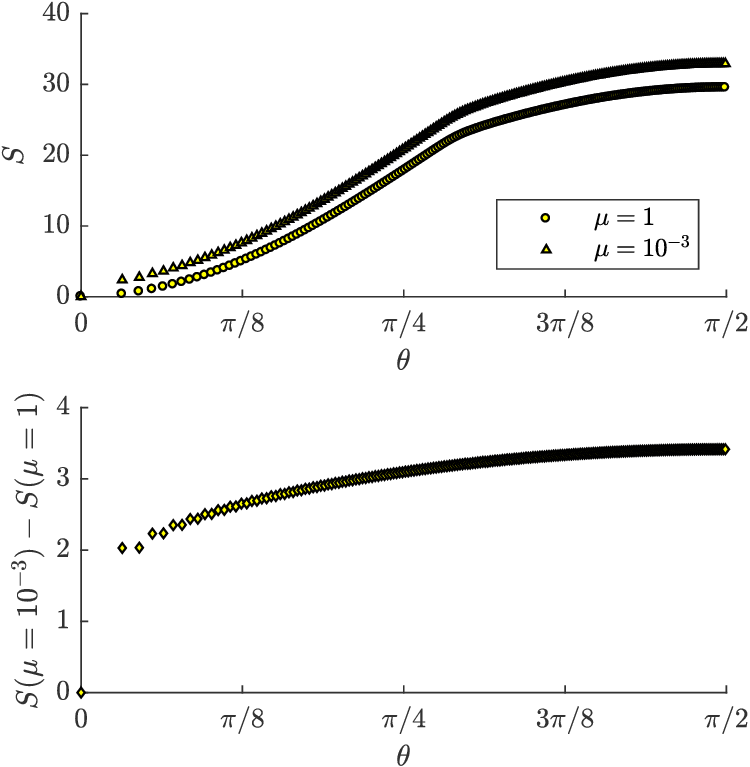}
    \caption{$ n/N=4/5$}
    \label{S_vs_theta_beta4over5_mu1mu1E-3}
    \end{subfigure}
    \caption{Entanglement entropies with $\mu=1$ and $\mu=10^{-3}$ 
and their difference as a function of $\theta$, for various values
of the UV cutoff parameter $n$.  $N=200$.}
    \label{S_beta1_mu1mu1E-3}
\end{figure}

Finally, we explore mass dependence of the entanglement
entropy.  In Figure \ref{vary_mu}, we show that entanglement
entropy with a UV cutoff has similar behaviour to that without
an imposed UV cutoff (which was discussed in \cite{Karczmarek:2013jca}),
being insensitive to changes in mass as long as $\mu < n$
and tending towards zero for $\mu >n$.
In Figure \ref{S_vs_theta_N200_mu}, we plot $S$ against $\theta$ for 
low and high masses. In the low mass case, the entanglement entropy
is nearly unchanged from $\mu=1$.  In the high mass case, as the
mass term dominates the kinetic energy terms, entanglement decreases,
still however retaining some of the qualitative features of the intermediate
mass case.
Figure \ref{S_beta1_mu1mu1E-3} further compares 
the $\mu=1$ and $\mu=10^{-3}$ cases.  We see that the main
effect of reducing the mass is a nearly-constant shift in
the entanglement entropy that is the result of an appearance
of a very-light mode at small mass.

\FloatBarrier 

\section{Mutual Information}
\label{sec:mutual}
Another quantity that behaves in an interesting way in
strongly coupled theories \cite{Karczmarek:2013xxa} is mutual information.
Mutual information is a useful quantity to study because it is
UV finite and---perhaps because of that--- is the same
on the fuzzy sphere as it is on the ordinary sphere 
\cite{Sabella-Garnier:2014fda}.  Given this, we compute
mutual information to validate our choice of method for
the assignment of degrees of freedom.

Mutual information is defined for a pair of regions.  Taking
those to be two polar caps  $C_1$ and $C_2$ centered at opposite 
poles of the sphere, mutual information is given by
\begin{align*}
    I = S(C_1)+S(C_2)-S(C_1 \cup C_2)~.
\end{align*}
To be able to compute it, we need to be able to assign degrees of freedom
(subspaces of $V^{(m)}_n$)
to caps $C_1$ and $C_2$ as well as their union $C_1 \cup C_2$.
One condition we must impose is that the linear subspace
associated with $C_1$ must be orthogonal to the linear
subspace associated with $C_2$; otherwise, we would
have a situation where two functions, despite of having support
on just one of two disjoint regions, have a non-zero overlap.
This can be easily and naturaly achieved with the operator
$Z_n^{(m)}$ defined in section \ref{sec:method}.  Given
angular sizes $\theta_1$ and $\theta_2$ 
of the two spherical caps $C_1$ and $C_2$ (with
$\theta_2$ measured from the south pole), we have
$z_1=(N-\half)\cos\theta_1$ and   $z_2=-(N-\half)\cos\theta_2$.
Degrees of freedom within $C_1$ are given by eigenvectors of
$Z_n^{(m)}$ with eigenvalues greater than $z_1$
and  degrees of freedom within $C_2$ are given by eigenvectors of
$Z_n^{(m)}$ with eigenvalues less than $z_2$.  Because
these eigenvectors are mutually orthogonal, $C_1 \cup C_2$
can be simply associated with the direct sum of these
two subspaces.

\begin{figure}
    \centering
    \includegraphics[width=\textwidth]{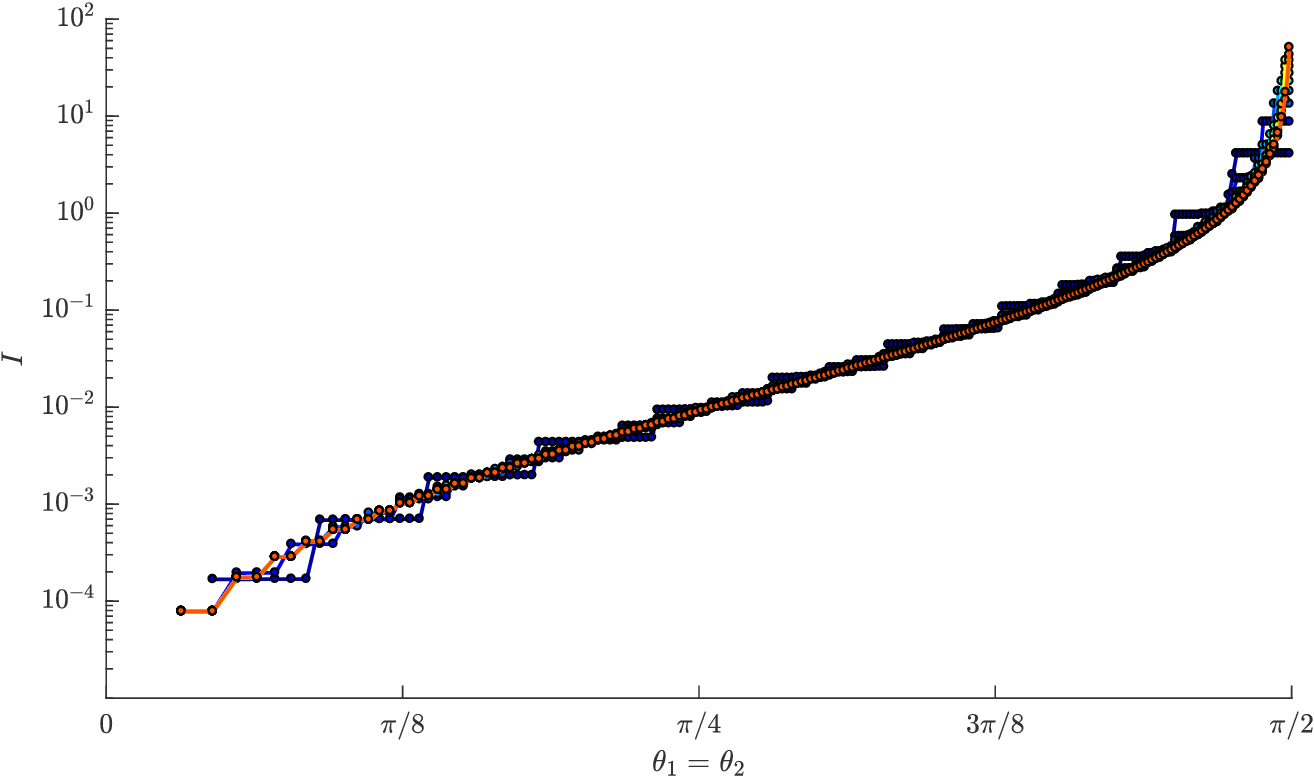}
    \caption{Mutual information $I(N,n;\theta)$ 
as a function of common angular size $\theta=\theta_1=\theta_2$ 
of two spherical caps centered at opposite poles of the sphere. 
$N=200$ and $\mu=1$. Colors ranging from dark blue to 
redish orange represent UV cutoff parameter $n=20,40,\ldots,200$.
}
    \label{I_vs_theta}
\end{figure}

For simplicity, we compute mutual information $I(N,n;\theta)$
when the two polar caps $C_1$ and $C_2$ 
have the same angular size $\theta$.  
Following our notation for entanglement entropy,
we denote mutual information on a commutative sphere with
$I(\infty,n;\theta)$.
In \cite{Sabella-Garnier:2014fda} it was shown that
$I(n,n;\theta)$ is independent of $n$ and that $I(n,n;\theta)=I(\infty,n;\theta)$. 
Given this, we would expect that $I(N,n;\theta)$ would be independent
of $n$ over the entire range of $N\in[n,\infty)$.  This is
indeed what is seen in Figure \ref{I_vs_theta}:
mutual information seems unaffected by the cutoff, except for
some artifacts having to do with discretization of the angle $\theta$,
which should go away at large $N$.  
This fact is further evidence that our prescription for
identifying degrees of freedom associated with spherical cap
regions works.  Mutual information turns out to be a nontrivial
test for the validity of degree of freedom assignment.
In the reminder of the paper, we discuss different
approaches to this assignment and explain why they appear to be inadequate.

In section \ref{sec:method} we defined a nearly-projection operator
$P^{(m)}_{\theta,n}:=O^{(m)}_n P^{(m)}_{\theta}  (O^{(m)}_n)^T$.
We now define\footnote{Omitting indices at this point
to reduce clutter.} $P_1:=O^{(m)}_n P^{(m)}_{\mathrm{north},\theta_1}  
(O^{(m)}_n)^T$
and $P_2:=O^{(m)}_n P^{(m)}_{\mathrm{south},\theta_2}  (O^{(m)}_n)^T$
and try to associate eigenvectors of $P_1$ and $P_2$
with eigenvalues greater than half with degrees of freedom of the
corresponding spherical cap.
Unfortunately,
while the associated subspaces are linearly
independent vector spaces,\footnote{This is easy 
to show.  Let $v$ be any normalized linear combination of eigenvectors
of $P_1$ with an eigenvalue greater than half.  Then,
$v^T P_1 v > \half$.  Since $\mathrm{Id}_{V_m}-P_{\theta_1}-P_{\theta_2}$ 
is obviously a positive semidefinite operator,
$\mathrm{Id}_{V_m^n}-P_1-P_2$ is also a positive
semidefinite operator.  Therefore, $v^T P_2 v < \half$ and $v$
cannot be a linear combination of eigenvectors
of $P_2$ with an eigenvalue greater than half.}
they are not guaranteed to be orthogonal.
In addition to re-assigning degrees of freedom slightly
to fix this, we also need to make sure that the vector
space associated with $C_1 \cup C_2$ is the direct
sum of vector spaces associated with $C_1$ and $C_2$.
This requirement introduces an ambiguity, because there
are degrees of freedom that do not appear to correspond to $C_1$
or $C_2$, but which could reasonably be thought of as belonging to $C_1\cup C_2.$\footnote{
Consider that there might exist a normalized vector $v$ in $V_n^{(m)}$
such that $v^T P_1 v < \half$ and $v^T P_2 v < \half$, but
$v^T (P_1 +  P_2) v > \half$.}
We can choose to associate  a direct
sum of vector spaces associated with $C_1$ and $C_2$
as the vector space associated with $C_1 \cup C_2$.
We call this the `exclusive' method as it appears to
include the smallest possible number of degrees of freedom with $C_1 \cup C_2$.
An alternative is an `inclusive' method, where we
first fix the vector space associated with $C_1 \cup C_2$
as the span of eigenvectors of 
$P_1 + P_2$ with eigenvalues greater than half, and then
decide how to break this vector space into a sum of two pieces 
associated with $C_1$ and $C_2$ respectively.  
This second method includes more degrees of freedom in $C_1 \cup C_2$
than the first.  
In either of the two cases, it is necessary to decide how to break
a vector space into two pieces associated with either spherical
cap.  We do this by examining the eigenvectors of $P_1-P_2$
(restricted to the space associated with $C_1 \cup C_2$). Those
associated with positive eigenvalues we assign to cap $C_1$ and
those associated with negative eigenvalues we assign to cap $C_2$.
Notice that if there is no UV cutoff imposed ($n=N$), $P_1$, $P_2$
and $P_1+P_2$ are projection operators, none
of the above ambiguities arise and both methods reduce to
that using $Z^{(m)}_n$.

Not only are the assignment methods based on $P_1$ and $P_2$
much more complicated than the one based on $Z^{(m)}_n$,
using them yields different results for mutual information.
Artifacts appear in mutual information $I(N,n;\theta)$ for
$\theta$ larger than the transition $\theta_\mathrm{trans}$.
Those artifacts are UV finite in the exclusive assignment method 
and grow like $N$ in the inclusive assignment method.
It would be interesting to understand further why out of
several
supperficially similar methods for assigning degrees of
freedom, only one leads to sensible mutual information.


\section*{Acknowledgments}

This work was completed with support from the
Natural Sciences and Engineering Council
of Canada (NSERC), grant SAPIN-2016-00032.

\bibliographystyle{JHEP}
\bibliography{with-entanglement}

\end{document}